\documentclass[letterpaper,preprint,notoc]{JHEP} 
\usepackage{axodraw}

\newcommand\fverb{\setbox\pippobox=\hbox\bgroup\verb}
\newcommand\fverbdo{\egroup\medskip\noindent%
			\fbox{\unhbox\pippobox}\ }
\newcommand\fverbit{\egroup\item[\fbox{\unhbox\pippobox}]}
\newbox\pippobox

\def\ksl{\not{\hbox{\kern-2.3pt $k$}}}
\def\lsim{\buildrel < \over {_\sim}}
\def\gsim{\buildrel > \over {_\sim}}

\def\e{\epsilon}

\def\oneloop{{\rm 1\! -\! loop}}
\def\twoloop{{\rm 2\! -\! loop}}

\def\Ord{{\cal O}}
\def\cm{{\cal M}}

\def\Re{{\rm Re}}

\def\MSbar{\overline{\rm MS}}

\def\li#1{{\mathop{\rm Li}\nolimits}_#1}
\def\ggtogg{\gamma\gamma \to \gamma\gamma}
\def\fig#1{fig.~{\ref{#1}}}
\def\Fig#1{Figure~{\ref{#1}}}
\def\eqn#1{eq.~(\ref{#1})}
\def\Eqn#1{Equation~(\ref{#1})}

\newskip\humongous \humongous=0pt plus 1000pt minus 1000pt

\newif\ifdtup

\newcounter{eqnumber}[section]

\preprint{hep-ph/0109079 \\     SLAC--PUB--8974\\
          UCLA/01/TEP/18\\
          September, 2001}

\title{QCD and QED Corrections to Light-by-Light Scattering}

\author{Z. Bern,\thanks{Research supported by the US Department of 
Energy under grant DE-FG03-91ER40662.} 
\ A. De Freitas$^*$ \\
Department of Physics and Astronomy \\
UCLA, Los Angeles, CA 90095-1547}

\author{L. Dixon\thanks{Research supported by the US Department of 
Energy under grant DE-AC03-76SF00515.}
\\
Stanford Linear Accelerator Center\\
Stanford University\\
Stanford, CA 94309}

\author{A. Ghinculov$^*$ and H.L. Wong$^*$\\
Department of Physics and Astronomy \\
UCLA, Los Angeles, CA 90095-1547}

\abstract{
We present the QCD and QED corrections to the fermion-loop contributions
to light-by-light scattering, $\ggtogg$, in the ultrarelativistic limit 
where the kinematic invariants are much larger than the masses of the 
charged fermions.
}


\dedicated{\centerline{Submitted to JHEP}}

\begin{document}

\renewcommand{\thefootnote}{\arabic{footnote}}
\setcounter{footnote}{0}


\section{Introduction}
\label{IntroSection}

Light-by-light scattering is one of the most fundamental processes in QED.
Theoretically, it proceeds at leading order, $\Ord(\alpha^4)$, via
one-loop box diagrams containing charged particles.  At center-of-mass
energies $\sqrt{s}$ far below the mass of the electron, the process is
described by the Euler-Heisenberg effective
Lagrangian~\cite{EulerHeisenberg}, and the cross section rises rapidly
with energy, $\sigma \propto s^3/m_e^8$.  The cross section peaks at
$\sqrt{s} \approx 3 m_e$~\cite{KarplusNeuman}, then begins to fall
rapidly, $\sigma \propto 1/s$ at fixed angles for $s \gg
m_e^2$~\cite{Akhiezer}.  At still higher energies, similar thresholds are
crossed for the muon, tau, and light quarks -- or rather, the light
hadrons.  (See \fig{OneLoopSigmaFigure}.)
The final significant Standard Model thresholds reached are
those of the $W$ boson~\cite{JT,GPRone} and the top quark.

The direct experimental evidence for $\ggtogg$ scattering is still scant,
particularly for energies above a few GeV.  At optical (electron volt)
energies, at least two experiments have been proposed to detect the
Euler-Heisenberg interaction via birefringence of the vacuum using laser
photons in a magnetic field~\cite{EHExpt}.  The Crystal Ball experiment at
the SPEAR storage ring had some unpublished evidence for $\ggtogg$
scattering at $\sqrt{s}$ of order several MeV, after subtracting beam-off
$\gamma$ ray backgrounds~\cite{Godfrey}.  More extensive data are
available for the process of Delbr\"uck scattering, in which two of the
four photons are supplied by the Coulomb field of a nucleus.  Several
experiments have studied this process, for incident photon energies
ranging from a few MeV to about a GeV.  Higher order Coulomb ($Z\alpha$)
corrections are often important in the comparison with
theory~\cite{Delbruck}.  Light-by-light scattering via an electron loop is
also tested quite precisely, if indirectly, by the measurement of
the anomalous magnetic moment of the electron~\cite{electrongm2},
as well as that of the muon~\cite{BNLgm2} (the latter is even
sensitive to muonic and hadronic light-by-light scattering).

\FIGURE{{\epsfxsize12cm \epsfbox{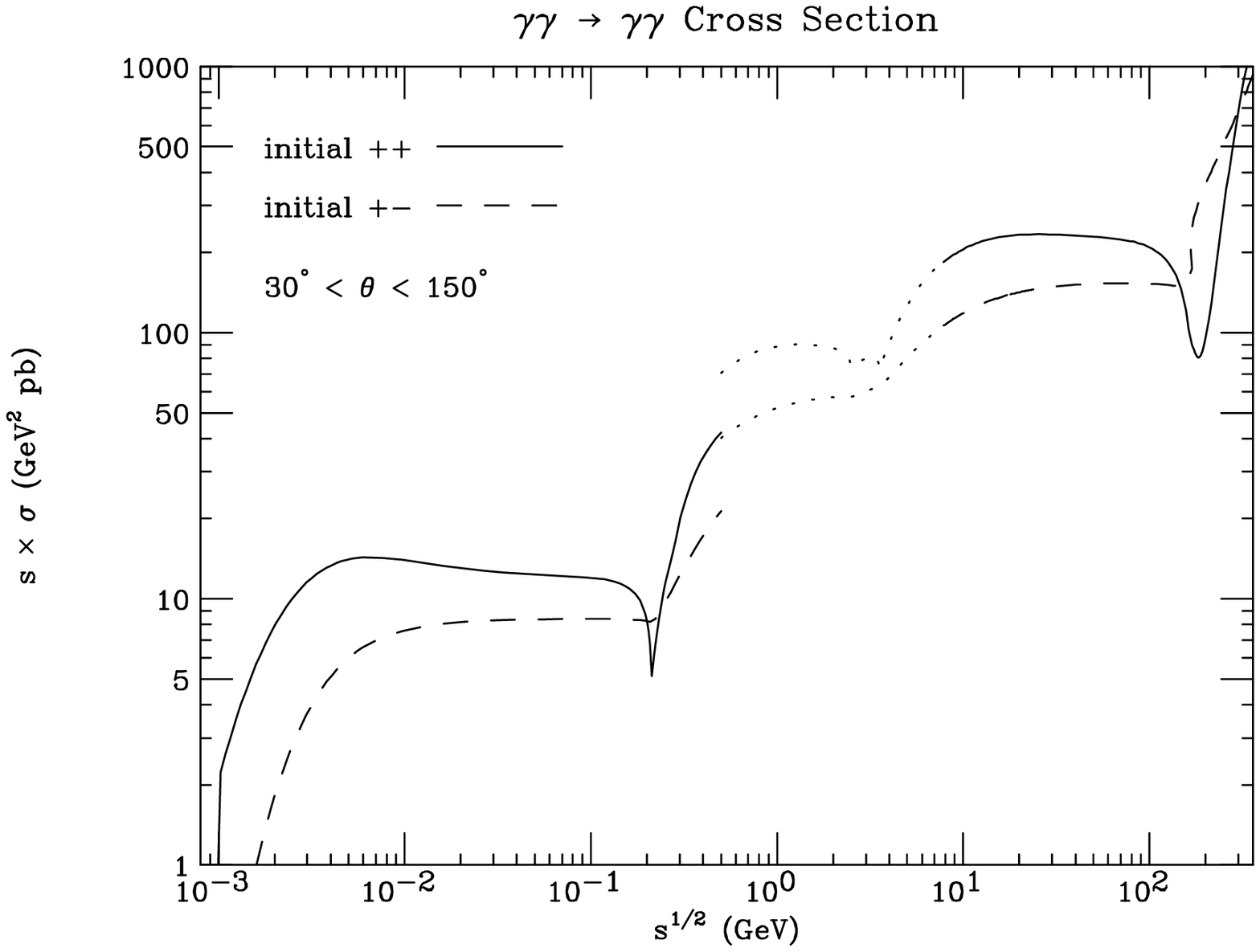}}
      \caption{
The leading order (one loop) cross section for light-by-light scattering,
for intermediate energies, showing the light fermion thresholds. 
The cross section has been integrated over a range of center-of-mass 
scattering angles, $30^\circ < \theta < 150^\circ$, for the two 
independent choices of initial photon helicities (assuming CP invariance), 
$++$ (solid) and $+-$ (dashed).  We have multiplied $\sigma$ by $s$ merely 
to compress the dynamic range.  We set $\alpha=1/137.036$.  
For $\sqrt{s} < 0.5$ GeV we omit all quark loops but include the charged pion
and kaon loops.   For $\sqrt{s} > 0.5$ GeV we use the quark loops, with
$m_u = m_d = 5$ MeV, $m_s = 100$ MeV, $m_c = 1.25$ GeV, $m_b = 4.2$ GeV, 
$m_t = 175$ GeV, and $m_W = 80.42$ GeV.  The hadronic contribution to 
the region from 0.5 to 8 GeV is not computed reliably by the quark boxes; 
hence that region is shown dotted.  We also omit the hadronic resonance 
contributions, $\pi^0$, $\eta$, etc.
} \label{OneLoopSigmaFigure} }

At much higher energies, the small size of the Standard Model
light-by-light scattering cross section provides a potential window to new
physics.  By backscattering a laser pulse off an intense, high energy 
electron beam~\cite{GammaBackscatter}, it
is possible to create $\gamma\gamma$ collisions with $\sqrt{s}$ of order
100--1000 GeV, high levels of initial state polarization, and
luminosities of order tens of fb$^{-1}$ per year.  
It has been shown that the $\ggtogg$ process at high energies is a sensitive
probe~\cite{Hooman} of theories with large extra dimensions~\cite{ADD},
for example.  Despite this interest in light-by-light scattering, the
process has to date only been calculated to leading order (one loop),
except in the low energy limit, where the two-loop corrections to the
Euler-Heisenberg Lagrangian are known~\cite{TwoLoopEulerHeisenberg}.

In this paper, we present the two-loop corrections
to $\ggtogg$ in the ultra-relativistic regime where
the kinematic invariants $s,t,u$ are much greater than the (squared) charged
fermion masses.  This regime is relevant for two ranges of center of mass
energy in the Standard Model:
\begin{itemize}
\item For $m_c \ll \sqrt{s} \lsim 2 m_W$ (and neglecting the tiny
bottom quark contribution), the QCD corrections from attaching 
a gluon line to the quark box are most important.  These give rise to
finite $\Ord(\alpha^4\alpha_s)$ corrections to the cross section.
\item For $m_e \ll \sqrt{s} \lsim 2 m_\mu \approx 2 m_\pi$, the dominant
corrections are QED corrections from attaching a photon line to the
electron box.  There are also QED corrections from inserting a fermion loop
onto an external leg; however, these are cancelled completely if the 
theory is renormalized at zero momentum transfer, i.e. by using 
$\alpha \equiv \alpha(0) = 1/137.036\ldots$ as the coupling constant.  
Then the two-loop QED corrections become identical to those of QCD, up to 
an overall constant.
\end{itemize}

The Feynman diagrams for $\ggtogg$ at two loops are a small subset of
those required for gluon-gluon scattering, $gg \to gg$.  The
interference of these two loop amplitudes with the $gg \to gg$ tree
amplitudes is an essential ingredient for obtaining the
next-to-next-to-leading order QCD corrections to jet production at
hadron colliders.  This interference was recently evaluated in a tour
de force calculation by Glover, Oleari and
Tejeda-Yeomans~\cite{GOTYgggg}.  In the light-by-light case, the tree
amplitudes vanish.  Thus the {\it next}-to-leading order corrections
require a different interference, of two-loop amplitudes with one-loop
amplitudes.  Instead of evaluating this interference directly, we have
computed the two-loop $\ggtogg$ amplitudes in a helicity basis.  Thus
polarized as well as unpolarized cross sections can be obtained,
information which is quite useful in the photon case, because of the
high initial state photon polarizations that are possible with
backscattering.

Rather than calculating the Feynman diagrams for the two-loop $\ggtogg$ 
helicity amplitudes, we employed a unitarity- or cut-based 
technique~\cite{CutBased,BRY,AllplusTwo} to generate the required loop 
momentum integrals.  These integrals were then evaluated using techniques
recently developed~\cite{IntegralsSV,IntegralsAGO,%
IntegralsTalks} to handle double box and related integrals where all 
internal lines are massless.  The restriction to massless internal lines
limits the validity of our results to the ultrarelativistic region where
all kinematic invariants are much greater than the relevant fermion masses.

Remarkably, some of the helicity amplitudes stay quite simple, even 
at two loops.  Some of this simplicity can be understood via unitarity 
and supersymmetry Ward identities~\cite{SWI}.

This paper is organized as follows.  In section 2 we present 
analytic results for the two-loop $\ggtogg$ helicity amplitudes.
(The more complicated formulae are relegated to an appendix.)  
In section 3 we give numerical results for the QCD and QED corrections.  
In section 4 we present our conclusions.


\section{The Two-Loop Amplitudes}
\label{AmplitudesSection}

The one-loop Feynman diagrams for photon-photon scattering via a
charged fermion consist of six box diagrams, which are related by 
permutations of the external photons to the diagram shown in 
\fig{DiagramsFigure}a.  At two loops, there are $6 \cdot 10 = 60$ 
nonvanishing Feynman diagrams, corresponding to connecting any two sides 
of the box with a photon propagator (or a gluon propagator, in the case
that the charged fermion is a quark).  Up to permutations, these
diagrams are depicted in \fig{DiagramsFigure}b.  Diagrams consisting of
two separate charged fermion triangles connected by a single photon or
gluon, as illustrated in \fig{DiagramsFigure}c, vanish by Furry's
theorem~\cite{Furry} or simple group theory.


\FIGURE{
\begin{picture}(350,240)(0,0)
\Text(-5,210)[c]{(a)}
\Text(10,175)[r]{1} \Text(10,234)[r]{2} 
\Text(75,234)[l]{3} \Text(75,175)[l]{4} 
\ArrowLine(29,192)(29,217) \Line(29,217)(54,217) 
\Line(54,217)(54,192) \Line(54,192)(29,192) 
\Photon(29,192)(12,175){2.5}{3} \Photon(29,217)(12,234){2.5}{3} 
\Photon(54,217)(71,234){2.5}{3} \Photon(54,192)(71,175){2.5}{3}  
\Text(-5,125)[c]{(b)}
\Text(10,90)[r]{1} \Text(10,149)[r]{2} 
\Text(75,149)[l]{3} \Text(75,90)[l]{4} 
\ArrowLine(23,101)(23,138) \Line(23,138)(60,138) 
\Line(60,138)(60,101) \Line(60,101)(23,101) 
\Photon(23,101)(12,90){2.5}{2} \Photon(23,138)(12,149){2.5}{2} 
\Photon(60,138)(71,149){2.5}{2} \Photon(60,101)(71,90){2.5}{2}  
\Gluon(41.5,101)(41.5,138){2.5}{4}
\Text(110,90)[r]{1} \Text(110,149)[r]{2} 
\Text(175,149)[l]{3} \Text(175,90)[l]{4} 
\ArrowLine(123,101)(123,138) \Line(123,138)(160,138) 
\Line(160,138)(160,101) \Line(160,101)(123,101) 
\Photon(123,101)(112,90){2.5}{2} \Photon(123,138)(112,149){2.5}{2} 
\Photon(160,138)(171,149){2.5}{2} \Photon(160,101)(171,90){2.5}{2}  
\Gluon(160,119.5)(141.5,138){2.5}{3}
\Text(210,90)[r]{1} \Text(210,149)[r]{2} 
\Text(275,149)[l]{3} \Text(275,90)[l]{4} 
\ArrowLine(223,101)(223,138) \Line(223,138)(260,138) 
\Line(260,138)(260,101) \Line(260,101)(223,101) 
\Photon(223,101)(212,90){2.5}{2} \Photon(223,138)(212,149){2.5}{2} 
\Photon(260,138)(271,149){2.5}{2} \Photon(260,101)(271,90){2.5}{2}  
\GlueArc(241.5,138)(10,180,360){2.5}{3}
\Text(-5,35)[c]{(c)}
\ArrowLine(29,22)(29,47) \Line(29,47)(51,34.5) \Line(51,34.5)(29,22)
\ArrowLine(97,47)(97,22) \Line(97,22)(75,34.5) \Line(75,34.5)(97,47)
\Photon(29,22)(12,5){2.5}{3} \Photon(12,64)(29,47){2.5}{3}
\Photon(114,5)(97,22){2.5}{3} \Photon(97,47)(114,64){2.5}{3}
\Gluon(51,34.5)(75,34.5){2.5}{3}
\Text(195,35)[c]{(d)}
\ArrowLine(229,22)(229,47) \Line(229,47)(254,47) 
\Line(254,47)(254,22) \Line(254,22)(229,22) 
\Photon(229,22)(221,14){2.5}{1.5} 
\ArrowArc(218,11)(4.2426,0,360)
\Photon(215,8)(207,0){2.5}{1.5} 
\Photon(229,47)(212,64){2.5}{3} 
\Photon(254,47)(271,64){2.5}{3} \Photon(254,22)(271,5){2.5}{3}  
\end{picture}
\caption{
Feynman diagrams for light-by-light scattering via a charged
fermion loop. (a) One of the six box diagrams contributing at one loop;
the remaining diagrams are obtained by permuting the external
photons.  (b) The three types of diagrams contributing at two loops; the
rest of the 60 diagrams are obtained by permutations.  
The curly line can be either a photon or a gluon, depending on 
whether the QED or QCD correction is being computed.
(c) These diagrams vanish by Furry's theorem or simple group theory.
(d) These diagrams could contribute to the QED correction.  However, 
they are precisely cancelled by conventional on-shell renormalization.}
\label{DiagramsFigure} }

We did not evaluate the Feynman diagrams directly.  Instead we
computed the unitarity cuts in various channels, working to all orders
in the dimensional regularization parameter $\epsilon = (4-D)/2$.
Essentially we followed the approach first employed at two loops for
the special cases of $N=4$ supersymmetric amplitudes~\cite{BRY} and 
the pure gluon four-point amplitude with all plus
helicities~\cite{AllplusTwo}.  These amplitudes were simple enough
that a compact expression for the integrand could be given.  The
fermion loop contributions with all plus helicities are about as
simple~\cite{TLS}.  However, for the generic helicity configuration, the 
integrands become quite complicated.  

We have used general integral reduction algorithms developed for the
all-massless planar four-point topologies~\cite{IntegralsSV,IntegralsAGO},
in order to reduce the loop integrals to a minimal basis of master
integrals.  Some mild extensions of these techniques are required in order
to incorporate polarization vectors for photons of definite
helicity~\cite{ggggpaper}.  We then expand the master integrals in a
Laurent series in $\e$, which begins at order $1/\e^4$.  Many of the
master integral Laurent expansions quoted in
refs.~\cite{IntegralsSV,IntegralsAGO} are in terms of Nielsen
functions~\cite{NielsenRef}, usually denoted by $S_{n,p}(x)$, with $n+p
\leq 4$.  However, using various identities~\cite{NielsenIds} the results
can be expressed in terms of the polylogarithms~\cite{Lewin},
\begin{eqnarray}
\li{n}(x) &=& \sum_{i=1}^\infty { x^i \over i^n }
          = \int_0^x {dt \over t} \li{{n-1}}(t),\nonumber \\
\li2(x) &=& -\int_0^x {dt \over t} \ln(1-t) \,, 
\label{PolyLogDef}
\end{eqnarray}
with $n=2,3,4$.

It is reassuring that all of the poles in $\e$ cancel for each helicity
amplitude in the QCD case.  In QCD, two loops is the first order at which
$\alpha_s$ appears in the $\ggtogg$ amplitude; therefore there can be no
ultraviolet divergence.  Any infrared divergence would have to be
cancelled by real gluon radiation; but the process $\ggtogg g$ is
forbidden by group theory.
  
In addition to infrared finiteness, the calculational framework was also
tested by gauge invariance: replacing a photon polarization vector by its
momentum vector produces a vanishing result. Finally, the same computer
programs for evaluating the cuts and reducing the integrals were also
used to compute the fermion-loop contribution to the two-loop $gg \to gg$
amplitudes in the helicity formalism and the 't Hooft-Veltman
dimensional regularization scheme~\cite{ggggpaper}.
The interference of the two-loop $gg \to gg$ helicity amplitudes with the
tree amplitudes, after summing over all external helicities and colors
and accounting for the different scheme used, is in complete agreement
with the calculation using conventional dimensional
regularization~\cite{GOTYgggg}.

The QED case requires exactly the same set of two-loop diagrams as in QCD,
up to an overall factor.  In addition, there are external fermion bubble
insertions of the form shown in \fig{DiagramsFigure}d.  In dimensional
regularization with massless fermions, these diagrams would vanish by
virtue of containing scale-free integrals.  This vanishing represents a
cancellation of ultraviolet and infrared divergences.  However, if one
renormalizes QED in the conventional on-shell scheme, to avoid infrared
divergences one should retain a fermion mass in the external bubbles.  Now
the bubble integral is nonzero and ultraviolet-divergent.  But this
divergence, and indeed the entire integral, is exactly cancelled by the
on-shell-scheme counterterm, precisely because the external leg is a real,
on-shell photon.

Hence the on-shell-renormalized two-loop QED-corrected amplitude is the
same as the two-loop QCD-corrected amplitude, up to overall coupling
constant factors.  In the on-shell scheme, the coupling constant should of
course be set to $\alpha \equiv \alpha(0) = 1/137.036\ldots$.  This value
should be used for all the QED couplings associated with the real,
external photons --- that is, for all couplings {\it except} the extra one
associated with the virtual photon in each two-loop graph.  The typical
virtuality of the extra photon is not zero, but of order $s$, assuming
that the kinematic invariants $s,t,u$ are all comparable in magnitude and 
much larger than the squared fermion mass, $m_f^2$.  Thus a running 
coupling $\alpha(\mu)$ with $\mu\approx\sqrt{s}$ should be used for the 
virtual photon insertion; whether it should be the $\MSbar$ running 
coupling, or that defined via the photon propagator at momentum transfer 
$\mu$, is theoretically indistinguishable at this order.  
For $m_e \ll \mu < m_\mu$, the latter coupling at one-loop order is
\begin{equation}
  \alpha(\mu)\ =\ { \alpha \over 
 1 - {\alpha\over3\pi} \Bigl[ \ln\Bigl({\mu^2 \over m_e^2}\Bigr) 
                              - {5\over3} \Bigr] } \,.
\label{RunningAlpha}
\end{equation}
For $\mu \gsim 2 m_\pi$, the running of $\alpha(\mu)$ receives 
hadronic corrections as well, which are best evaluated via a dispersion 
relation using the $e^+e^- \to $ hadrons data.  In any event, the precise
$\alpha(\mu)$ used makes very little difference, since it only appears in
an $\Ord(\alpha)$ correction to a process with a rather tiny cross section.

For the QCD corrections, we use the $\MSbar$ running 
coupling $\alpha_s(\mu)$, again with $\mu = \sqrt{s}$, keeping in mind 
that our calculation is only to leading order in $\alpha_s$.

We consider the process
\begin{equation}
 \gamma(k_1,\lambda_1) + \gamma(k_2,\lambda_2) 
\ \to \ \gamma(k_3,\lambda_3) + \gamma(k_4,\lambda_4),
\label{KinematicsDef}
\end{equation}
where $k_i$ and $\lambda_i$ are the photon momenta and helicities.
In terms of the center-of-mass energy $\sqrt{s}$ and scattering angle
$\theta$, the Mandelstam variables are $s = (k_1+k_2)^2$, 
$t = (k_1-k_4)^2 = -s/2 \times (1-\cos\theta)$, and 
$u = (k_1-k_3)^2 = -s/2 \times (1+\cos\theta)$, 
with $s>0$, $t<0$, $u<0$.
Parity, time-reversal invariance, and Bose symmetry imply that there are
only four independent helicity amplitudes, $M_{--++}$, $M_{-+++}$,
$M_{++++}$, and $M_{+--+}$.  Actually, crossing symmetry relates
$M_{++++}$ and $M_{+--+}$.  However, representing the two-loop amplitudes
in a crossing-symmetric fashion in terms of master integrals would lead to
more cumbersome formulae, so we shall present all four amplitudes instead.
We adopt the overall phase convention of refs.~\cite{JT,GPR}.  Then
formula (9) in ref.~\cite{GPR} can be applied to our helicity amplitudes
in order to obtain the $\ggtogg$ differential cross section for generic
circular and transverse photon polarizations (Stokes parameters).

The one-loop helicity amplitudes due to a fermion of
mass $m_f$ in the loop take a very simple form in the ultra-relativistic 
limit, $\{s,t,u\} \gg m_f^2$~\cite{Akhiezer,KarplusNeuman,GvdB}.  We write
\begin{equation}
\cm^\oneloop_{\lambda_1\lambda_2\lambda_3\lambda_4}
= 8 \, N Q^4 \, \alpha^2 \,
    M_{\lambda_1\lambda_2\lambda_3\lambda_4}^{(1)} \,,
\label{OneLoopDecomp}
\end{equation}
where $N$ is the fermion color factor (3 for quarks, 1 for leptons),
and $Q$ is the fermion charge in units of $e$.  The functions
$M_{\lambda_1\lambda_2\lambda_3\lambda_4}^{(1)}$ are given by
\begin{eqnarray}
M_{--++}^{(1)} &=& 
   1 
\,, \nonumber \\
M_{-+++}^{(1)} &=& 
   1, 
 \nonumber \\
M_{++++}^{(1)} &=& 
- {1\over2} {t^2+u^2\over s^2} 
                  \Bigl[ \ln^2\Bigl({t\over u}\Bigr) + \pi^2 \Bigr]
               - {t-u\over s} \ln\Bigl({t\over u}\Bigr) - 1
\,,  \nonumber \\
M_{+--+}^{(1)} &=&
 - {1\over2} {t^2+s^2\over u^2} 
                  \ln^2\Bigl(-{t\over s}\Bigr)
               - {t-s\over u} \ln\Bigl(-{t\over s}\Bigr) - 1 \nonumber \\
&& \null \hskip 2 cm 
       - i \pi \biggl[ {t^2+s^2\over u^2} \ln\Bigl(-{t\over s}\Bigr)
                      + {t-s\over u} \biggr] 
\,.
\label{OneLoopFunctions}
\end{eqnarray}

The QCD- and QED-corrected two-loop amplitudes are
\begin{eqnarray}
\cm^{\twoloop,\,{\rm QCD}}_{\lambda_1\lambda_2\lambda_3\lambda_4}
&=& 4 \, (N^2-1) Q^4 \, \alpha^2 \, {\alpha_s(\mu) \over \pi}
    \, M_{\lambda_1\lambda_2\lambda_3\lambda_4}^{(2)} \,, \nonumber\\
\cm^{\twoloop,\,{\rm QED}}_{\lambda_1\lambda_2\lambda_3\lambda_4}
&=& 8 \, N Q^6 \, \alpha^2 \, {\alpha(\mu) \over \pi}
    \, M_{\lambda_1\lambda_2\lambda_3\lambda_4}^{(2)} \,,
\label{TwoLoopDecomp}
\end{eqnarray}
where the explicit values of the 
$M_{\lambda_1\lambda_2\lambda_3\lambda_4}^{(2)}$ 
are given in Appendix~\ref{AmplitudeAppendix}.

Note from \eqn{FppppSL}
that $M_{--++}$ remains remarkably simple even at two loops --- it is
just a constant, independent of $s$, $t$ and $u$!  This simplicity
is actually predictable.  The lack of an imaginary part for
$M_{--++}^{(2)}$ (as for $M_{--++}^{(1)}$ and $M_{-+++}^{(1)}$) can be
deduced by considering the possible unitarity cuts of the amplitude.  Let
$f$ be a massless fermion.  Then a supersymmetry Ward Identity~\cite{SWI}
shows that the tree-level amplitudes for $\gamma\gamma f \bar{f}$ and
$\gamma\gamma\gamma f \bar{f}$ vanish if all the photons have the same
helicity (when considered as outgoing particles).  The same result is true
if one (or more) photons are replaced by gluons with the same helicity.
Assigning opposite (outgoing) helicities to intermediate particles on
opposite sides of a cut, it is easy to see that all two- or three-particle
cuts vanish for the all-outgoing-plus-helicity two-loop amplitude
$M_{--++}^{(2)}$.
This argument assumes that a $D=4$ helicity assignment is valid for the
intermediate particles, a condition which is justified for the processes
considered here by the absence of infrared or ultraviolet divergences.  Thus
$M_{--++}^{(2)}$ must be a dimensionless rational function.  Crossing
symmetry implies that it is totally symmetric in $s,t,u$.  Imposing at
most single poles in $s,t,u$, and using $s+t+u=0$, one finds that 
$M_{--++}^{(2)}$ must actually be a constant.

\Eqn{FmpppSL} shows that $M_{-+++}^{(2)}$ is also fairly simple,
containing only logarithms, and not $\li2$, $\li3$ or $\li4$ functions.
Its two-particle cuts in $D=4$ are the product of two rational functions
(just like the cuts of a one-loop amplitude), which may partially account
for this simplicity; however, the three-particle cuts do not similarly
simplify at a glance.

Interestingly, up to the overall normalization these amplitudes are
identical to the subleading color contributions for the $gg \rightarrow
\gamma \gamma$ amplitudes presented in ref.~\cite{gggamgamPaper}.
(In that paper all particles are taken to be outgoing, which corresponds 
to flipping the helicity labels for particles 1 and 2.)
These amplitudes are relevant for improved estimates of the di-photon 
background to production of a light (mass $<$ 140 GeV) Higgs boson at 
the Large Hadron Collider~\cite{HiggsBkgdPaper}.

It is instructive to quote the values of the helicity amplitudes
in various limits.
The values of the one-loop functions for $90^\circ$ scattering 
($t=-s/2$, $u=-s/2$) are
\begin{eqnarray}
M_{--++}^{(1)}(90^\circ) &=& 1 \,, \nonumber\\
M_{-+++}^{(1)}(90^\circ) &=& 1 \,, \nonumber\\
M_{++++}^{(1)}(90^\circ) 
            &\approx&  -3.46740 \,, \nonumber\\
M_{+--+}^{(1)}(90^\circ) 
            &\approx& -0.12169 + 0.46574 \pi i\,,
\hbox{~~~~~~~~~~~~~~~~~~~~~~~~~~~~~~~~~~~~~~~~~~} 
\label{M1RightAngle}
\end{eqnarray}
while the two-loop functions are
\begin{eqnarray}
M_{--++}^{(2)}(90^\circ) &=& -1.5, \nonumber \\
M_{-+++}^{(2)}(90^\circ) 
      &\approx& 0.17770 - 0.23287 \pi i \,, \nonumber \\
M_{++++}^{(2)}(90^\circ) 
          &\approx& 1.24077 + 1.00717 \pi i \,, \nonumber \\
M_{+--+}^{(2)}(90^\circ)
                           &\approx& 0.01445 + 0.36840 \pi i \,.
\hbox{~~~~~~~~~~~~~~~~~~~~~~~~~~~~~~~~~~~~~~~~~~} 
\label{M2RightAngle}
\end{eqnarray}
Note that except for the first, the two-loop functions are
smaller than their one-loop counterparts.

In the small-angle limit $|t| \ll s$, the values of the one-loop amplitudes, 
up to corrections suppressed by powers of $t/s$, are
\begin{eqnarray}
M_{--++}^{(1)} &\sim& 
1
\,, \nonumber \\
M_{-+++}^{(1)} &\sim& 
1
\,, \nonumber \\
%
%
M_{++++}^{(1)} &\sim&
 - {1\over2} X^2 - X - {\pi^2 \over 2} - 1
 \,,\nonumber \\
M_{+--+}^{(1)} &\sim&
 - {1\over2} X^2 - X - 1 - i \pi ( X + 1 )
\,, 
\hbox{~~~~~~~~~~~~~~~~~~~~~~~~~~~~~~~~~~~~~~~~~~~~~~~~~~~~~~~~~~~~}\nonumber \\
M_{+-+-}^{(1)} &\sim&
 0
\,,\nonumber \\
\label{M1smallt}
\end{eqnarray}
while the two-loop amplitudes are,
\begin{eqnarray}
M_{--++}^{(2)} &\sim &
 -{3\over2} 
      \, ,  \nonumber \\
M_{-+++}^{(2)} &\sim & 
{1\over4} X^2 + {1\over2} X 
                     + {\pi^2\over8} - {1\over4}
         + i \pi \Bigl( {1\over4} X + {1\over4} \Bigr)
       \, , \nonumber \\
M_{++++}^{(2)} &\sim &
 - {1\over24} X^4 - {1\over12} X^3
           - \Bigl( {\pi^2\over12} + {1\over2} \Bigr) X^2
           - \Bigl( {5\over12} \pi^2 + {1\over2} \Bigr) X
           + {7\over360} \pi^4 - \zeta_3 - {7\over12} \pi^2 - {1\over4}
\nonumber \\ &&\null \hskip1cm 
         + i \pi \Bigl( {1\over2} X^2 + {1\over2} X
            + 2 \zeta_3 + {\pi^2\over6} - {13\over4} \Bigr)
     \,, \nonumber \\
M_{+--+}^{(2)} &\sim & 
- {1\over24} X^4 - {1\over12} X^3
           + \Bigl( {\pi^2\over6} - {1\over2} \Bigr) X^2
           + \Bigl( {5\over6} \pi^2 - {1\over2} \Bigr) X
           + {11\over180} \pi^4 - \zeta_3 + {5\over12} \pi^2 - {1\over4}
\nonumber \\ && \null \hskip1cm 
         + i \pi \Bigl( -{1\over6} X^3 - {3\over4} X^2 - {3\over2} X 
                      - 2 \zeta_3 + {11\over4} \Bigr)  
  \,, \nonumber \\
M_{+-+-}^{(2)} &\sim&
 -{1\over2} \, ,
\label{M2smallt}
\end{eqnarray}
where $X \equiv -\ln(-s/t)$.  Notice that the small-angle scattering
amplitudes where both photons flip their helicity have no logarithmic
enhancements at one- or two-loops; those with one helicity flip have
no logs at one loop but a factor of $X^2$ at two loops; and those with
no helicity flips pick up one factor of $X^2$ for each loop. 
For small angles these contributions are actually power
suppressed compared to $t$-channel vector exchange diagrams which first 
appear at three loops~\cite{LeadingLogs}.  

Next we give the differential cross sections in terms of the above
amplitudes.  The leading-order $\ggtogg$ unpolarized differential
cross section for a single fermion flavor is given by
\begin{equation}
 {d\sigma^{\rm LO} \over d\cos\theta} 
  = N^2 Q^8 \, {\alpha^4 \over 2 \pi \, s} \Bigl[
     | M_{--++}^{(1)} |^2 + 4 \, | M_{-+++}^{(1)} |^2 
   + | M_{++++}^{(1)} |^2 + | M_{+--+}^{(1)} |^2 + | M_{+-+-}^{(1)} |^2 
   \Bigr] \,,
\label{LeadingCrossSection}
\end{equation}
where 
\begin{equation}
M_{+-+-}^{(L)}(s,t,u) = M_{+--+}^{(L)}(s,u,t) \,.
\label{mppmBySymmetry}
\end{equation}
The QCD- and QED-corrected unpolarized differential cross sections for a
single ferm\-ion flavor are given by
\begin{eqnarray}
   {d\sigma^{\rm QCD} \over d\cos\theta} 
&=&  {d\sigma^{\rm LO} \over d\cos\theta}
 + {d\sigma^{\alpha_s} \over d\cos\theta} \,, 
\nonumber \\
   {d\sigma^{\rm QED} \over d\cos\theta} 
&=&  {d\sigma^{\rm LO} \over d\cos\theta}
 + {d\sigma^{\alpha} \over d\cos\theta} \,, 
\label{QCDCrossSection}
\end{eqnarray}
where
\begin{eqnarray}
 {d\sigma^{\alpha_s} \over d\cos\theta} 
  &= & N (N^2-1) Q^8 \, 
     {\alpha^4 \, \alpha_s(\mu) \over 2 \pi^2 \, s} I^{(2,1)} \,, \nonumber\\
 {d\sigma^{\alpha} \over d\cos\theta} 
  &=& N^2 Q^{10} \, 
     {\alpha^4 \, \alpha(\mu) \over \pi^2 \, s} I^{(2,1)} \,,
\label{alphasCrossSection}
\end{eqnarray}
and
\begin{eqnarray}
I^{(2,1)} &\equiv&
\Re \Bigl[
          M_{--++}^{(2)} M_{--++}^{(1)\ *}  
   + 4 \, M_{-+++}^{(2)} M_{-+++}^{(1)\ *}  
   +      M_{++++}^{(2)} M_{++++}^{(1)\ *}  
\nonumber \\ && \null \hskip 2 cm 
   +      M_{+--+}^{(2)} M_{+--+}^{(1)\ *}  
   +      M_{+-+-}^{(2)} M_{+-+-}^{(1)\ *}  \Bigr] \,. 
\label{I21}
\end{eqnarray}
The cross sections for circularly polarized photons can be 
constructed easily from \eqn{alphasCrossSection} by reweighting the 
products of $M$'s.  For arbitrary initial photon polarizations, linear
as well as circular, eqs.~(9)--(15) of ref.~\cite{GPR} can be used, 
after making the replacement (for the QCD case)
\begin{equation}
F_{\lambda_1\lambda_2\lambda_3\lambda_4} \to 
\cm_{\lambda_1\lambda_2\lambda_3\lambda_4}
= 8 \, N Q^4 \, \alpha^2 \, M_{\lambda_1\lambda_2\lambda_3\lambda_4}^{(1)}
+ 4 \, (N^2-1) Q^4 \, \alpha^2 \, {\alpha_s(\mu) \over \pi}
    \, M_{\lambda_1\lambda_2\lambda_3\lambda_4}^{(2)} \,.
\label{Freplace}
\end{equation}
(Note that our definitions of $t$ and $u$ are reversed with respect
to ref.~\cite{GPR}.)


\section{Numerical Results}

The QCD $K$ factor is conventionally defined as the ratio of
next-to-leading to leading-order cross sections,
\begin{equation}
 K \equiv  { d\sigma^{\rm QCD}/d\cos\theta 
      \over d\sigma^{\rm LO}/d\cos\theta } \,.
\label{KDef}
\end{equation}
In a region between quark thresholds, and below the $W$ mass, we can write
\begin{equation}
 K = 1 + {\alpha_s(\mu) \over \pi} 
     { \sum_i (N_i^2-1) Q_i^4 \over \sum_i N_i Q_i^4 } \, k,
\label{KEqn}
\end{equation}
where
\begin{equation}
 k \equiv  
{ \sum_{ \{ \lambda \} } \Re \Bigl[ M_{ \{ \lambda \} }^{(2)}
                                    M_{ \{ \lambda \} }^{(1)\ *} \Bigr]
  \over \sum_{ \{ \lambda \} } | M_{ \{ \lambda \} }^{(1)} |^2 } \,.
\label{kDef}
\end{equation}
The label $i$ runs over the number of quarks and leptons with masses
much less than $\sqrt{s}$; $N_i=3$ for quarks, $N_i=1$ for leptons;
and $Q_i$ is the fermion charge.  For energies between $m_b$ and
$m_W$, for example, $(\sum_i (N_i^2-1) Q_i^4 ) / ( \sum_i N_i Q_i^4 )
= 70/87$.  For energies above $m_W$, one should add the $W$ loop
contributions to $\cm^\oneloop_{\{ \lambda \} }$.  This dilutes the
QCD corrections considerably, since the $W$ loops quickly 
dominate~\cite{JT,GPRone} (see \fig{OneLoopSigmaFigure}), 
and they have no QCD corrections.

%
\FIGURE{ {\epsfxsize12cm \epsfbox{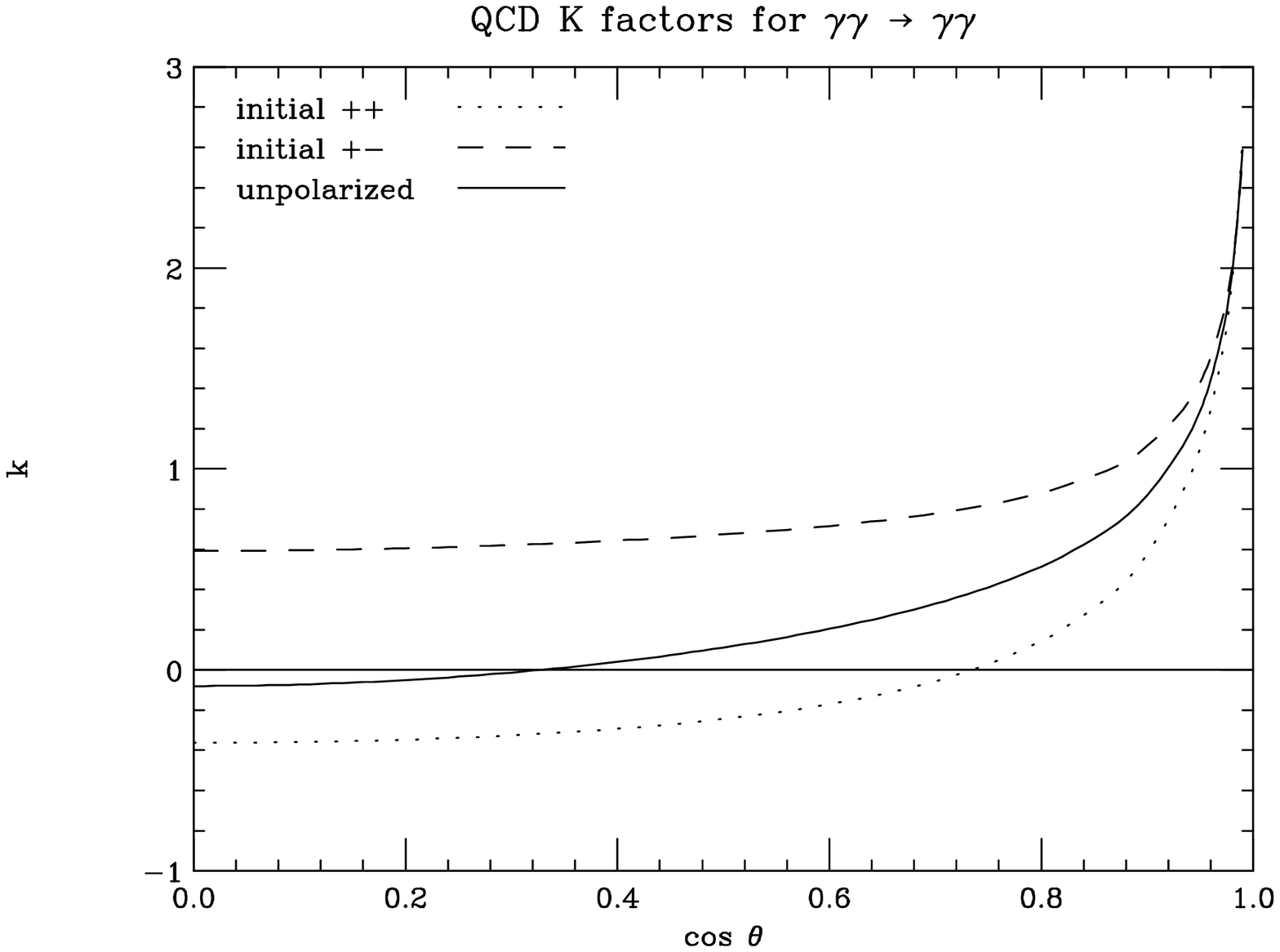}}
\caption{
Eqs.~(\ref{KEqn})--(\ref{klambdaDef}) express the QCD-correction $K$ factor
in terms of the quantities $k$ (unpolarized initial and final photons), 
$k_{++}$ and $k_{+-}$ (fully polarized initial photons).
Here $k$ (solid), $k_{++}$ (dotted) and $k_{+-}$ (dashed) are plotted as 
a function of $\cos\theta$.} \label{kpolFigure} }

In \fig{kpolFigure}, $k$ is plotted as a function of the
center-of-mass scattering angle $\theta$.  Due to Bose symmetry, only 
the forward region $\cos\theta > 0$ has to be plotted.
We also plot the corresponding curves for fully circularly polarized 
initial photons, but summing over final-state helicities,
\begin{equation}
 k_{\lambda_1\lambda_2} \equiv  
{ \sum_{\lambda_3,\lambda_4} 
  \Re \Bigl[ M_{ \{ \lambda \} }^{(2)} M_{ \{ \lambda \} }^{(1)\ *} \Bigr]
  \over \sum_{\lambda_3,\lambda_4} | M_{ \{ \lambda \} }^{(1)} |^2 } \,.
\label{klambdaDef}
\end{equation}
Taking into account CP invariance, there are two independent cases, 
$++$ and $+-$.  Although the QCD corrections are of order $\alpha_s/\pi$ for 
these two cases, they have opposite signs, and in the
unpolarized cross section there is a large cancellation.  The QCD
corrections to the unpolarized cross section are only of order 
$0.1 \times \alpha_s/\pi$ for central scattering angles, becoming as large
as $\alpha_s/\pi$ only for $|\cos\theta| \geq 0.95$.

\Fig{QCDCorrFigure} displays the QCD corrections to the cross section
in the 10--100 GeV region where quark boxes are important, and the
quarks can be taken to be approximately massless.  The QED corrections
can be neglected in this region as they are a factor of 8 or so
smaller.  The two independent initial helicity configurations are
shown, after integrating the differential cross section with an
acceptance cut of $30^\circ < \theta < 150^\circ$, and multiplying by
$s$ as in \fig{OneLoopSigmaFigure}.  The QCD corrections were computed
by multiplying the leading order one-loop result, which contains the
full fermion (and $W$) mass dependence, by the $K$ factor
formula~(\ref{KEqn}) evaluated for massless fermions in both numerator
and denominator (which are of course integrated separately over
$\theta$).  We can apply \eqn{KEqn} even across the $b$ quark
threshold, simply because the $b$ quark's small charge of $-1/3$ lends
it a tiny contribution to the $\ggtogg$ cross section.  (Nevertheless,
we interpolate $( \sum_i (N_i^2-1) Q_i^4 ) / (\sum_i N_i Q_i^4 )$ from
$272/345 \approx 0.788$ to $70/87 \approx 0.805$ across the
threshold.)  We see that the QCD corrections are numerically rather
small for central scattering angles in this entire energy range.  The
small size of the correction explicitly demonstrates the reliability
of the leading order prediction.

\FIGURE{
{\epsfxsize12cm \epsfbox{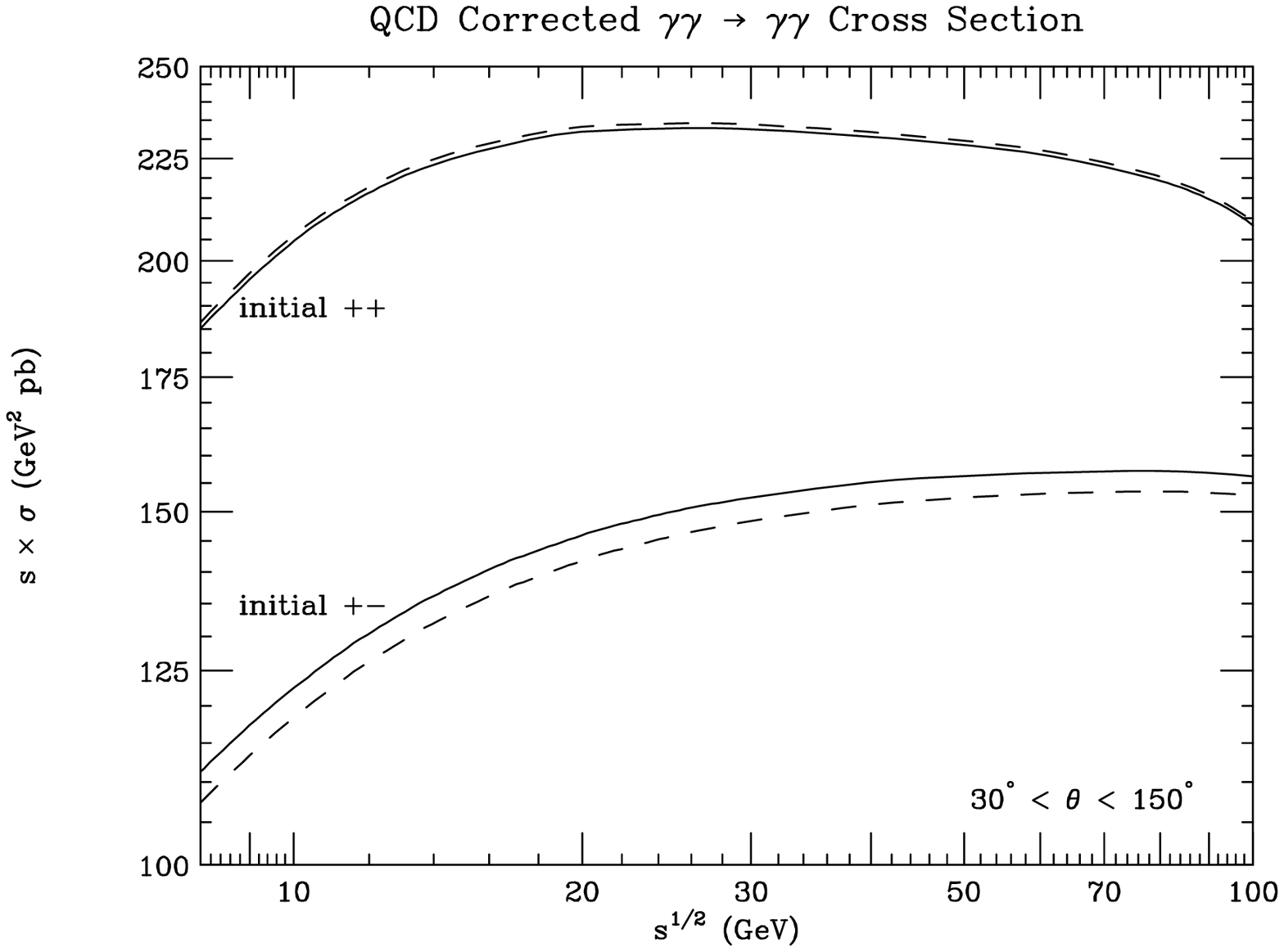}}
\caption{
QCD corrections to $\ggtogg$ in the intermediate energy region
8 GeV $< \sqrt{s} <$ 100 GeV where quark boxes are important,
for $\alpha_s(m_Z) = 0.118$, and renormalization scale $\mu=\sqrt{s}$.
Plotted are the results at leading order (dashed), and including the 
QCD corrections (solid), for the same quark masses used in 
\fig{OneLoopSigmaFigure}.}\label{QCDCorrFigure}}

The QED corrections can also be extracted from \fig{kpolFigure}.
In the region between $m_e$ and $m_\mu$, for example, the corrections
to the unpolarized cross section are
\begin{equation}
 { d\sigma^{\rm QED} \over d\cos\theta } =  
 { d\sigma^{\rm LO} \over  d\cos\theta } \times \biggl[ 1 
   + 2 {\alpha(\mu) \over \pi} \, k \biggr] \,.  
\label{QEDKfactor}
\end{equation}
For the same angular cuts ($30^\circ < \theta < 150^\circ$) as in the
QCD case, this amounts to only a $0.07\%$ decrease in the cross
section for $++$ initial helicities, and a $0.35\%$ increase for $+-$
initial helicities.  It will be a true challenge to measure the
$\ggtogg$ reaction to this level of precision.


\section{Conclusions}

In this paper we presented the two-loop QCD and QED corrections to 
light-by-light scattering by fermion loops in the ultrarelativistic limit
where all kinematic invariants are much greater than the relevant fermion 
masses.  These corrections reliably give the leading Standard Model 
corrections for most energies below 100 GeV.  The corrections are quite 
small numerically, showing that the leading order computations are robust.

Some of the helicity amplitudes remain quite simple even at two loops,
as a consequence of unitarity and a supersymmetry Ward identity for tree
amplitudes.

To extend our results to regions where the kinematic invariants are
comparable to the masses in the loops, the technology for computing
two-loop double box integrals should first be extended to include massive
internal lines, which seems feasible.  Probably the most important
application would then be to compute the electroweak corrections to the
$W$ box contribution to $\ggtogg$, since that contribution dominates at
high energies, where new physics contributions are most likely to be
found.


\acknowledgments
We thank Hooman Davoudiasl and George Gounaris for useful conversations.


\appendix
\section{Two-loop helicity Amplitudes}
\label{AmplitudeAppendix}

The explicit expressions for the two-loop amplitudes appearing in 
\eqn{TwoLoopDecomp} are
\begin{equation}
M_{--++}^{(2)} =
- {3\over 2 }\,, 
 \hskip 12. cm \null 
\label{FppppSL}
\end{equation}
\begin{eqnarray}
M_{-+++}^{(2)}  & = &  
    {1\over 8}  \biggl[ { x^2 + 1\over y^2 } ((X + i \pi)^2 + \pi^2 )
             + {1\over 2} (x^2 + y^2) ((X-Y)^2 + \pi^2)  \nonumber \\
&& \hskip 4 cm \null
             - 4  \biggl({1\over y} - x \biggr)  (X + i \pi) \biggr] 
+ \Bigl\{t \leftrightarrow u \Bigr\} \,,
 \hskip 2.8 cm \null
\label{FmpppSL}
\end{eqnarray}
\begin{eqnarray}
M_{++++}^{(2)} &=&  
 - 2 x^2 \biggl[ \li4(-x) + \li4(-y) 
  - (X + i\pi) \Bigl( \li3(-x) + \li3(-y) \Bigr)  \nonumber \\
&& \hskip1.0cm 
  + {1\over12} X^4 - {1\over3} X^3 Y + {\pi^2 \over 12} X Y 
  - {4 \over 90} \pi^4
  + i {\pi\over6} X \Bigl( X^2 - 3 X Y + \pi^2 \Bigr) \biggr] \nonumber \\
&& \null
-(x-y) \Bigl( \li4(-x/y) - {\pi^2\over6} \li2(-x) \Bigr) \nonumber\\ 
&& \null
 - x \biggl[ 2 \li3(-x) - \li3(-x/y) - 3 \zeta_3
      - 2 (X + i\pi) \li2(-x) \nonumber \\
&& \hskip1.0cm 
      + (X-Y) ( \li2(-x/y) + X^2 ) 
      + {1\over 12} ( 5 (X-Y) + 18 i \pi) ((X-Y)^2 + \pi^2) \nonumber \\
&& \hskip1.0cm 
      - {2\over 3}  X  (X^2 + \pi^2) - i \pi (Y^2 + \pi^2) \biggr]
\nonumber \\
&& \null
  + {1\over 4} { 1 - 2x^2 \over y^2 } ((X + i\pi)^2 + \pi^2)
  - {1\over 8} ( 2 x y + 3 ) ((X-Y)^2 + \pi^2)
  + { \pi^2 \over 12} \nonumber \\
&& \null
  + \biggl( {1\over 2 y} + x \biggr) (X + i\pi) - {1\over 4} 
                    +  \Bigl\{ t \leftrightarrow u \Bigr\} \,,
\label{FmmppSL}
\end{eqnarray}
\begin{eqnarray}
M_{+--+}^{(2)} &=&  
- 2 {x^2+1 \over y^2} \biggl[ 
       \li4(-x/y) - \li4(-y) 
    + {1\over2} (X - 2 Y - i\pi) ( \li3(-x) - \zeta_3 ) \nonumber \\
&& \hskip2.0cm
    + {1\over 24} ( X^4 + 2 i \pi X^3 - 4 X Y^3 + Y^4 
              + 2 \pi^2 Y^2 ) + {7\over 360} \pi^4 \biggr] \nonumber \\
&& \null
 - 2 {x-1\over y} \biggl[ \li4(-x) - \zeta_4
       - {1\over 2} (X + i\pi) (\li3(-x) - \zeta_3 )  \nonumber \\
&& \hskip2.0cm
       + {\pi^2\over 6} \Bigl( \li2(-x) - {\pi^2\over6} 
                              - {1\over2} X^2 \Bigr) 
       - {1\over 48} X^4 \biggr]  \nonumber \\
&& \null
 + \biggl(2 {x\over y} - 1\biggr) 
    \biggl[ \li3(-x) - (X + i\pi) \li2(-x) 
            + \zeta_3 - {1\over 6} X^3
            - {\pi^2\over 3} (X + Y) \biggr]  \nonumber \\
&& \null
 + 2 \biggl(2 {x\over y} + 1 \biggr) 
     \biggl[ \li3(-y) + (Y + i\pi) \li2(-x) - \zeta_3 
     + {1\over 4} X ( 2 Y^2 + \pi^2 )   \nonumber \\
&& \hskip2.5cm
          - {1\over 8} X^2 (X + 3 i\pi) \biggr]
     - {1\over 4} ( 2 x^2 - y^2 ) ((X-Y)^2 + \pi^2)  \nonumber \\
&& \null 
     - {1\over 4} \Bigl(3 + 2 {x\over y^2} \Bigr) ((X + i\pi)^2 + \pi^2)
     - {2-y^2 \over 4 x^2} ((Y + i\pi)^2 + \pi^2)  
     + {\pi^2\over 6}  \nonumber \\
&& \null
     + {1\over 2} ( 2 x + y^2 ) \biggl[ {1\over y} (X + i\pi) 
                                     + {1\over x} (Y + i\pi) \biggr] 
                 - {1\over 2} \,.
\label{FmpmpSL}
\end{eqnarray}
Here
\begin{equation}
x \equiv {t\over s} \,, \quad y \equiv {u\over s} \,, \quad
X \equiv \ln\Bigl(-{t\over s}\Bigr) \,, \quad
Y \equiv \ln\Bigl(-{u\over s}\Bigr) \,.
\label{XYdef}
\end{equation}
%


\end{document}